\documentclass[ aip,
 amsmath,amssymb,
 reprint,twocolumn,
]{revtex4-1}

\usepackage{graphicx}
\usepackage{dcolumn}
\usepackage{bm}

\usepackage[utf8]{inputenc}
\usepackage[T1]{fontenc}
\usepackage{mathptmx}
\usepackage{etoolbox}
\usepackage{xcolor}

\definecolor{mediumtealblue}{rgb}{0.0, 0.33, 0.71}

\makeatother
\begin{document}

\preprint{AIP/123-QED}

\title{Relations between normal state nonreciprocal transport and the superconducting diode effect in the trivial and topological phases}

\author{Georg Angehrn}
\thanks{These authors contributed equally}
\affiliation{Department of Physics, University of Basel, Klingelbergstrasse 82, CH-4056 Basel, Switzerland}

\author{Henry F. Legg}
\thanks{These authors contributed equally}
\affiliation{Department of Physics, University of Basel, Klingelbergstrasse 82, CH-4056 Basel, Switzerland}

\author{Daniel Loss}
\affiliation{Department of Physics, University of Basel, Klingelbergstrasse 82, CH-4056 Basel, Switzerland}
\author{Jelena Klinovaja}
\affiliation{Department of Physics, University of Basel, Klingelbergstrasse 82, CH-4056 Basel, Switzerland}

\date{\today}

\begin{abstract}
Nonreciprocal transport effects can occur in the normal state of conductors and in superconductors when both inversion and time-reversal symmetry are broken. Here, we consider systems where magnetochiral anisotropy (MCA) of the energy spectrum due to an externally applied magnetic field results in a rectification effect in the normal state and a superconducting (SC) diode effect when the system is proximitised by a superconductor. Focussing on nanowire systems, we obtain analytic expressions for both normal state rectification and SC diode effects that reveal the commonalities -- as well as differences -- between these two phenomena. Furthermore, we consider the nanowire brought into an (almost) helical state in the normal phase or a topological superconducting phase when proximitised. In both cases this reveals that the topology of the system considerably modifies its nonreciprocal transport properties. Our results provide new insights into how to determine the origin of nonreciprocal effects and further evince the strong connection of nonreciprocal transport with the topological properties of a system.
\end{abstract}

\maketitle

\section{\label{secIntro} Introduction} A diode is a device in which the resistance depends on the direction of current flow and is a fundamental element of most modern electronics. A diode effect requires a non-reciprocal resistance, $R$, due to a current, $I$, such that $R(+I)\neq R(-I)$. In other words, it requires the current to be rectified. To achieve such an effect it is necessary that inversion symmetry and time-reversal symmetry are both broken simultaneously. In most cases time-reversal symmetry is broken by dissipation and the required inversion symmetry breaking is achieved extrinsically, e.g., by forming a $pn$-junction.

In contrast, however, it is also possible for a nonreciprocal resistance to arise as an intrinsic property of a material, e.g., due to the band structure. For instance, applying an external magnetic field can result in a magnetochiral anisotropy (MCA) of the band structure that also results in a diode effect. \cite{Rikken2001, Rikken2005, Pop2014, Ideue2017, Yokouchi2017, Tokura2018, He2018,Wang2022,Legg2022MCA,Dantas2023} In particular, MCA results in a resistance that is proportional to the current, $I$, itself, such that $R= R_0(1+\gamma B I)$, where 
$R_0$ is the reciprocal resistance, $B$ is the magnetic field strength, and $\gamma$ the MCA rectification coefficient.\cite{Ideue2017}

Whilst the above diode effects occur in the normal state, it has also recently been discovered that it is possible to have a diode effect in superconductors.\cite{Qin2017,Ando2020,Baumgartner2021} In this case a superconducting material or Josephson junction exhibits a critical current, $I_c^\pm$, that is dependent on the direction of current flow, such that e.g. $I^+_c > |I^-_c|$, where $\pm$ indicates the direction of current flow.\cite{Yokoyama2014,Ando2020,Baumgartner2021,Lyu2021,Ilic2022,Wu2022,Hou2022,Souto2022,He2022,Daido2022,Noah2022,Pal2022,Legg2022,Baumgartner2022,Bauriedl2022,Diez2023,Nadeem2023,Ciaccia2023,Legg2023,Costa2023,schrade2023,he2023,Gupta2023,Souto2023,Costa2023micro,zhao2023,chiles2023,Trahms2023,Scheurer2024,Zhang2024,Valentini2024} These effects are known as the {\it superconducting (SC) diode} and {\it Josephson diode} effect, respectively. This difference in critical current enables a diode effect because there exists a range of currents -- in the above example $|I^-_c|<|I|<I^+_c$ -- that experiences the zero-resistance of a superconductor in one direction but has a finite resistance for current flow in the opposite direction.\cite{Ando2020,Baumgartner2021}

Several mechanisms have been proposed to produce SC diode effects. As in normal state diodes, broken inversion symmetry -- either explicitly or intrinsically -- and broken time-reversal symmetry -- either explicitly by a magnetic field~\cite{Daido2022,Noah2022,Legg2022} or by some other mechanism~\cite{Wu2022,Souto2022,Hou2022,Scheurer2024,Trahms2023,zhao2023,chiles2023} -- are required. In particular, it has been shown that MCA of the underlying energy spectrum in a superconductor can also result in a SC diode or a Josephson diode effect.~\cite{Baumgartner2021,Legg2022,Baumgartner2022,Legg2023}

\begin{figure}[t]
\includegraphics[width=1\linewidth]{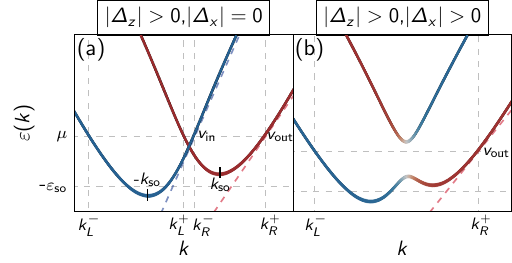}
\caption{{\it Energy spectrum}: Schematic normal state band structure, $\varepsilon(k)$, of a one-dimensional system with spin-orbit interaction (SOI), i.e., a nanowire. Inversion symmetry breaking plays a crucial role in both the case of normal state rectification and the SC diode effect. In particular, a difference in Fermi velocity for the inner and outer branches of the band structure ($v_{\rm in}\neq v_{\rm out}$) directly results in an SC diode effect. The difference in velocity also stems from the difference in band curvature for the inner and outer Fermi points, which is what results in a normal state rectification. Both the normal state rectification and SC diode effect are substantially modified when a magnetic field component opens a gap at $k=0$, as in (b), resulting in an (almost) helical state of the normal state and a topological superconducting state in the proximitised system. }
\label{fig1}
\end{figure}

Although normal state rectification and the SC diode effect can both stem from MCA of the energy spectrum,~\cite{Legg2022MCA,Legg2022,Bauriedl2022,Baumgartner2022} the relationships between the two effects have not been explored. Understanding how MCA of the normal state and superconducting phase relate to each other can provide evidence a given SC diode effect is due to intrinsic broken inversion symmetry and could provide a roadmap to producing better SC diodes based on MCA which has significant potential technological applications.~\cite{schrade2023,Souto2023,Nadeem2023,Valentini2024,Ciacca2024} Furthermore, such a relationship could provide a measure of the reduction of spin-orbit interaction (SOI) due to metallisation effects~\cite{reeg2018,Legg2022metal} in hybrid SC devices, which is an essential ingredient for realising topological superconductivity~\cite{Prada2020} and controlling Andreev spin-qubits in such devices.~\cite{Spethmann2022} Finally, it has been shown that SC diode effects can be altered dramatically in the topological superconducting phase,\cite{Rosdahl2018,Legg2022,Legg2023} which is still not fully understood, and further insights into the connections with topology could be gained from the normal state rectification effect.

In this paper, we investigate systems with MCA of the energy spectrum and consider relations between the normal state rectification effect and SC diode effect when proximitised by a superconductor. Focussing on nanowire systems,~\cite{Prada2020,Laubscher2021} we obtain analytic expressions for both normal state rectification and SC diode effects that reveal the links in how both phenomena precisely rely on inversion symmetry breaking of the bulk band structure. We also investigate the case where the nanowire is brought into an (almost) helical state in the normal phase or a topological superconducting phase when proximitised. This reveals, in both the normal state and proximitised systems, that the topology of the system considerably modifies the nonreciprocal transport response. Interestingly, the normal state rectification is generically enhanced in the (almost) helical phase, whereas the SC diode effect is substantially reduced. We conclude with a discussion of the relations between these two effects and how this can potentially be leveraged to achieve better SC diodes, as well as provides insights into the origins of nonreciprocal effects.

\section{Nonreciprocal transport in the normal state of nanowires}
We will first investigate normal state nonreciprocal transport in a (quasi) one-dimensional nanowire. In particular, we focus on the diffusive regime with a scattering rate $1/\tau$ as, for instance, considered in recent experiments on topological insulator nanowires.\cite{Legg2022MCA} For simplicity we consider only a single subband of a one-dimensional band structure. As such, we assume that the system is shorter than the localization length and do not take into account other conductivity channels, e.g., in the bulk of the TI nanowire. Although sufficient for our purposes to understand relations between normal state rectification and the SC diode effect, these contributions can still be important in real experimental systems since they affect $\gamma$, the rectification coefficient.

\subsection{Calculation of nonlinear conductivity}
The normal state of a nanowire in the presence of a magnetic field is described by the following Hamiltonian,~\cite{Legg2022}
\begin{equation}
\label{hk}
h_{k} = \xi_{k}\sigma_{0} + (\alpha_{k} + \Delta_{z})\sigma_{3}+ \Delta_{x}\sigma_{1}-\mu,
\end{equation}
where $\xi_{k}=\xi_{-k}$ (to be defined later),  $\Delta_{z}$ ($\Delta_x$) is the Zeeman energy due to a magnetic field parallel (perpendicular) to the SOI vector, i.e. direction, of strength $\alpha_{k} = -\alpha_{-k}$,
defining the quantisation axis, and $\sigma_{i}$ denote the Pauli matrices. The eigenenergies  are  described by an energy dispersion $\varepsilon_k^{(s)}$, where $k$ is the momentum along the nanowire and $s$ the band index. We will first consider the case of a magnetic field that is parallel to the SOI  vector [Fig.~1(a)] and then later consider the case where an additional component of magnetic field perpendicular to the SOI vector results in an (almost) helical pair of states at Fermi levels within the resulting gap [Fig.~1(b)].

To calculate the current, we expand the Fermi distribution function in powers of electric field, $E$, applied along the nanowire, such that $f_l(k)\propto E^l $, and then solve the Boltzmann equation in the constant relaxation time approximation (as in, e.g., Refs.~\citenum{Ideue2017,Legg2022MCA,Dantas2023}) such that the $l$th order current  contribution is given by 
\begin{align}\label{nonlincond}
j_l=\sigma_l E^l
=\frac{e}{2 \pi}\left(\frac{e \tau E}{\hbar}\right)^l \sum_{s=\pm,\, i\in\{L,R\}} \operatorname{sgn}\left[v_{k_{i}^{s}}^{s}\right] \mathcal{V}_{l, k_{i}^{s}}^{s},
\end{align}
where $e<0$ is the electron charge, $\hbar \mathcal{V}_{l, k}^{s}=\partial_k^l \epsilon_k^{(s)}$, $\hbar v_k^{s}=\partial_k \epsilon_k^{(s)}$, and $k_{i}^{s}$ is the right (R) or left (L) Fermi wavevector associated with the band $s=\pm$ [blue/red in Fig.~1(a)].  The full current at all orders of electric field is given by $j=\sum_l j_l$. 

\subsection{Rectification due to magnetic field parallel to SOI vector}

We now consider the case of a magnetic field parallel to the SOI vector (see Fig.~1) and expand around each band minimum to cubic order in $\delta k_s=k-s k_{\rm S O}$ with $s=\pm$ the band index and where $s k_{\rm S O}$ is the momentum corresponding to the band minima in the absence of magnetic fields [see Fig.~\ref{fig1}(a)], such that
\begin{equation}
\label{eig:en}
\varepsilon^{(s)}_k \approx -\varepsilon_{\rm S O}-\mu-s \Delta_z+\frac{\alpha_m}{2}\left(\delta k_{s}\right)^2+\frac{s \beta}{6}\left(\delta k_{s}\right)^3,
\end{equation}
where the quadratic coefficient $\alpha_m>0$ is related to the band mass  and the coefficient $\beta$ is related to cubic SOI.
Here, $\mu$ is the chemical potential and $\Delta_z=g \mu_B B_z/2$ is the Zeeman energy due to the magnetic field $B_z$ along the SOI axis with $\mu_B$ the Bohr magneton and $g$ the g-factor. Note, the chemical potential, $\mu$, is measured from the band crossing point at $k=0$ and $-\epsilon_{\rm so}$ is the energy of the band minima, both in the absence of a magnetic fields.  We will see that this is the minimal order expansion required to obtain a finite nonreciprocal effect.
We also note that we assume that $\beta$ is not too large ($\beta k_{\rm S O} \ll \alpha_m$) such that the quadratic expansion is already a good approximation of the bands and cubic corrections are small. In this case, we could ignore the spurious Fermi points at large momentum arising from the cubic order, alternatively,  an additional quartic term can remove these spurious crossings.~\cite{Luethi2023}

Using Eq.~\eqref{nonlincond} to calculate the second-order conductivity, i.e. the leading order nonreciprocal component, at temperature $T=0$, we obtain
\begin{align}
\sigma_2^{\rm n}&= \frac{e}{2\pi}\left(\frac{e\tau}{\hbar}\right)^{2}\frac{\beta}{\hbar}\left(k_{R}^{+}-k_{R}^{-}-k_{L}^{+}+k_{L}^{-}\right)\\
&\approx\frac{e}{\pi}\left(\frac{e \tau}{\hbar}\right)^2 \frac{\beta}{\hbar^2} \Delta_z \left(\frac{1}{v_{\text {in }}}+\frac{1}{v_{\text {out }}}\right),\nonumber
\end{align}
where $v_{\rm in}$ ($v_{\rm out}$) are the Fermi velocities of the inner (outer) branches of the spectrum in the absence of a magnetic field [see Fig.~\ref{fig1}(a)]. In the second line the approximation is to leading order in $\beta$ and $\Delta_z$, which are treated perturbatively, such that $k_{R}^{+}+k_{L}^{-}\approx 2\Delta_z/\hbar v_{\rm out}$ and $k_{R}^{-}+k_{L}^{+}\approx -2\Delta_z/\hbar v_{\rm in}$. We note that, in general, $v_{\rm out}\neq v_{\rm in}$ if $\beta\neq 0$, which is required to achieve a finite $\sigma_2^n$. A significant cubic contribution, $\beta$, arises, for instance, in germanium nanowires\cite{Dantas2023} or topological insulator (TI) nanowires.\cite{Legg2022MCA}  We note that in the absence of the cubic term in our expansion of the energy spectrum, $\beta=0$, the nonreciprocal coefficient will vanish unless higher orders of momentum in Eq.~\eqref{eig:en} are included. In other words, a purely quadratic dispersion, even with broken inversion symmetry, is not sufficient to produce a finite nonreciprocal conductivity.~\cite{Legg2022MCA,Dantas2023} In the absence of magnetic fields, due to the presence of time-reversal symmetry, as expected, $\sigma_2^n$ vanishes.

\subsection{Rectification in the (almost) helical state}
We now consider the case where an additional component of magnetic field parallel to the nanowire (perpendicular to the SOI vector) opens a partial gap at small momentum. The chemical potential $\mu$ can be placed inside this partial gap and the system hosts a pair of (almost) helical states~\cite{streda2003,Braunecker2010,Klinovaja2011,Rainis2014} [see Fig.~1(b)].
Similar to above, using Eq.~\eqref{nonlincond} to calculate the second order conductivity gives us
\begin{equation}
\sigma_2^{\rm h}=\frac{e}{2\pi}\left(\frac{e\tau_h}{\hbar}\right)^{2}\frac{\beta}{\hbar}\left(k_{R}^{+}+k_{L}^{-}\right)\approx\frac{e}{\pi}\left(\frac{e \tau_h}{\hbar}\right)^2 \frac{\beta}{\hbar^2}\frac{\Delta_z }{v_{\text {out}}},
\end{equation}
where $\tau_h$ is the scattering time in the helical state.
As such we find the simple relationship between the second order conductivities in both cases
\begin{equation}
\sigma_2^{\rm h}/\sigma_2^{\rm n} \approx \left(\frac{\tau_h}{\tau}\right)^2 \frac{v_{\rm in}}{v_{\rm in}+v_{\rm out}},
\end{equation}
where $v_{\rm in}$ and $v_{\rm out}$ are, as above, the Fermi velocities without an applied magnetic field.

The nonreciprocal conductivity can be either reduced or enhanced, depending on the relative ratios of $v_{\rm in}/v_{\rm out}$ and $\tau_h/\tau$. However, backscattering is reduced in the (almost) helical state and so we generically expect that the scattering time will be much longer for this case, $\tau_h\gg \tau$. Therefore, the increase of scattering time expected in an (almost) helical state and the fact it appears as the square of the $\tau_h/\tau$, likely results in an increase of $\sigma_2$.

\section{Superconducting diode effect in proximitised nanowires}
We now turn to the SC diode effect in nanowires.~\cite{Legg2022,Gupta2023,he2023,Zhang2024} We will assume that the superconductivity in the nanowire is induced by the proximity effect and therefore no self-consistent treatment of the pairing potential is required. Such a setup was considered numerically in Ref.~\citenum{Legg2022}, where it was shown that a change of sign and dramatic reduction in the SC diode efficiency occurred for values of Zeeman energy, $\Delta_x$, when the topological superconducting phase transition had occurred. Here, we establish analytically the origin of the SC diode effect and the connection to the topological superconducting phase transition. These insights will allow us to see the common features with the normal state rectification effect discussed above. 

\begin{figure}[t]
\includegraphics[width=1\linewidth]{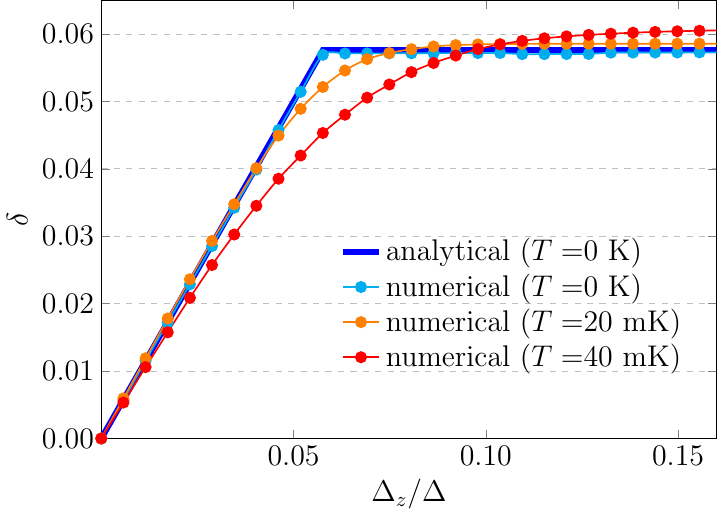}
\centering
\caption{{\it Superconducting diode efficiency}: The diode efficiency initially increases linearly with Zeeman energy before saturating at a Zeeman field that is set by the difference in Fermi velocities of the inner and outer branches, see Eq.~\eqref{delta}. The linearised approximation utilised for analytic calculations (dark blue line) is a good approximation for the diode efficiency at low temperatures.  For the numerical solution, the full energy spectrum of $\mathcal{H}_{k,q}$ [see Eq.~\eqref{hsc}] is used to find the maxima and minima of the current [see Eq.~\eqref{currentden}] at the indicated finite temperature. We use a $h_k$, as defined in Eq.~\eqref{hk} with $\xi_{k}=\frac{\hbar}{2m_{0}}k^{2} + Ck^{4}$ and $\alpha_k=\alpha_{\rm SO}k$. Here $m_{0}$ is the band mass, $C$ the quartic contribution that results in $v_{\rm out}\neq v_{\rm in}$ necessary for a finite SC diode effect, and $\alpha_{\rm SO}$ the linear spin-orbit coefficient.
Parameters: $m_{0}=0.015\text{ }m_{e}$, $C=2\cdot10^{5}\text{ meV nm}^{4}$, $\alpha_{SO}=0.65\text{ eV }$\AA, $\mu=0.4\text{ meV}$, $\Delta=0.2\text{ meV}$, $\Delta_x=0$. 
}
\label{delta1D}
\end{figure}

\subsection{Calculation of supercurrent}

When brought into proximity with a superconductor a pairing potential, $\Delta$, is induced in the nanowire.  Furthermore, the presence of a current through the nanowire can be described by a finite Cooper pair momentum, $q$, in the nanowire.\cite{Noah2022,He2022,Daido2022,Legg2022} As such the full Hamiltonian in the presence of a supercurrent and pairing potential is given by $H_{q}=\frac{1}{2}\int dk \psi_k^\dagger \mathcal{H}_{k,q} \psi_k$, where
\begin{equation}\label{hsc}
\mathcal{H}_{k,q} = \begin{pmatrix} h_{{k}+{q}/2} & -i\sigma_{2}\Delta \\ i\sigma_{2}\Delta & -h_{-{k}+{q}/2}^{\ast} \end{pmatrix},
\end{equation}
and, as in Eq.~\eqref{hk}, $h_k$ is the normal state Hamiltonian.
Here, $\psi_k^{\dagger}=(c_{k+\frac{q}{2} \uparrow}^{\dagger},\ c_{k+\frac{q}{2} \downarrow}^{\dagger},\ c_{-k+\frac{q}{2} \uparrow},\ c_{-k+\frac{q}{2} \downarrow})$, where the creation (annihilation) operator $c_{k \sigma}^\dagger$ ($c_{k \sigma}$) acts on an electron with momentum $k$ and spin $\sigma$. Throughout we will assume that $\Delta\geq0$ is real.

For now we focus only on the case when the magnetic field is parallel to the SOI vector. To calculate the current we follow the same approach as in Ref.~\citenum{Legg2022} and to make analytic progress we linearise the normal state spectrum in the absence of magnetic fields, which will be taken later into account perturbatively. Furthermore, we assume that superconducting pairing only occurs between equivalent Fermi points, i.e., only between outer and only between inner points, such that they can be treated separately.~\cite{Klinovaja2012}  In this case the Hamiltonian of each pair  of Fermi points can be described by 
\begin{equation}
h_{\eta, k}= \hbar v_\eta (- \eta  \sigma_3 k - k_\eta^0)+\Delta_z \sigma_3,
\end{equation}
where $\eta=+1 (-1)$ indicates the outer (inner) Fermi points, such that the $k_{-1}^0$ ($k_{1}^0$) is the Fermi wave number of the inner (outer) point of the $+$ band at zero magnetic field and velocities $v_{-1}=v_{\rm in}$ and $v_{1}=v_{\rm out}$ are as defined above. In Fig.~\ref{delta1D}, we show that this approximation captures well  the SC diode behavior at low temperatures. For each pair of Fermi points, $\eta$, the two eigenenergies are given by~\cite{Legg2022} (see the Supplemental Material (SM)~[\onlinecite{SI}])
\begin{equation}
E_{n,\eta}(\Delta,q,k)=\sqrt{\Delta ^2+\hbar^2 v_\eta^2 ( \eta k +n k^0_\eta)^2}+n(-\hbar v_\eta \eta q/2 +\Delta_z), \label{energy}
\end{equation}
where $n=\pm$.  Here, we chose two branches that correspond to positive energies in the absence of supercurrents and magnetic fields. The remaining two branches are related to $E_{n,\eta}(\Delta,q,k)$ by particle-hole symmetry and are given by $-E_{-n,\eta}(\Delta,q,-k)$.
As in previous studies,~\cite{Daido2022,Legg2022} the current can then be obtained from the free energy density (see the SM~[\onlinecite{SI}] for details) such that 
\begin{align}
\label{currentden}
j(q)=&\frac{e}{\pi \hbar} \int dk \sum_{\eta,n=\pm}\bigg[\tanh \left(\beta E_{n,\eta}(0, q,k)/2\right) \partial_{q}E_{n,\eta}(0,q,k)  \nonumber \\
&-\tanh \left(\beta E_{n,\eta}(\Delta, q,k)/2\right) \partial_{q} E_{n,\eta}(\Delta,q,k)\bigg],
\end{align}
where $\beta=(k_B T)^{-1}$, with $T$ being the temperature and $k_B$ the Boltzmann constant.

\subsection{Ground state Cooper pair momentum for magnetic field parallel to SOI.}
\begin{figure}[t]
\includegraphics[width=1\linewidth]{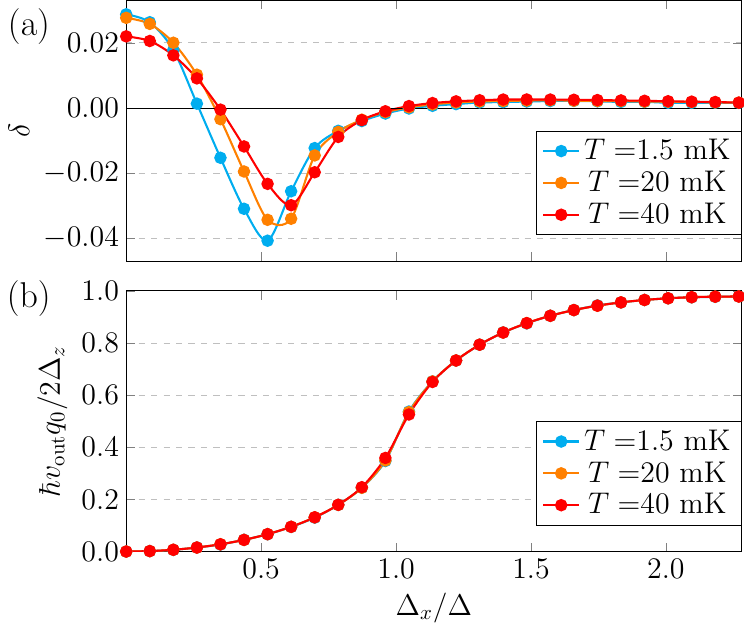}
\centering
\caption{{\it Superconducting diode efficiency of the topological phase}: Numerical calculation of  (a) the diode coefficient $\delta$ and (b) the ground-state Cooper pair momentum $q_{0}$ as a function of the Zeeman energy $\Delta_{x}$ resulting from a magnetic field parallel to the nanowire. Deep within the topological regime ($\Delta_x \gg \Delta$), we find that the diode efficiency is considerably suppressed by the presence of a finite pairing momentum $q_0$ in the ground state of the superconductor. We find $q_0$ is not strongly affected by small finite temperature. We use a normal state Hamiltonian, $h_k$, as in Eq.~\eqref{hk} with $\xi_{k}=\frac{\hbar}{2m_{0}}k^{2} + Ck^{4}$ and $\alpha_k=\alpha_{\rm SO}k$. Here, $m_{0}$ is the band mass, $C$ the quartic contribution that results in $v_{\rm out}\neq v_{\rm in}$ necessary for a finite SC diode effect, and $\alpha_{\rm SO}$ the linear SOI coefficient. Parameters: $m_{0}=0.015\text{ }m_{e}$, $C=2\cdot10^{5}\text{ meV nm}^{4}$, $\alpha_{\rm SO}=0.65\text{ eV }$\AA, $\Delta=0.2\text{ meV}$, $\mu=0\text{ meV}$, $\Delta_{z}=0.006\text{ meV}$.} 
\label{topo}
\end{figure}

When subject to a magnetic field, the ground state superconducting condensate could develop a finite Cooper pair momentum.~\cite{Kinnunen2018,Daido2022,Noah2022} If there is a finite gap and the energies $E_{n,\eta}(\Delta, q,k) >0$ for all $k$, in the zero-temperature limit, the current is given by 
\begin{align}
 j(q)= &\frac{e}{2\pi }\sum_{\eta,n=\pm} n \eta v_{\eta}\int d{k} \;  (1-\text{sgn}\{ E_{n,\eta}(0,q,k)\}) \nonumber\\
&=\frac{e}{\pi } \sum_{\eta}  v_\eta \left(q - \frac{2\eta \Delta_z}{\hbar v_\eta } \right) = \frac{e}{\pi } q \left(v_{\rm in}+v_{\rm out}\right), \label{jq} 
\end{align}
where the first line follows from the fact that, at zero temperature, the terms in Eq.~\eqref{currentden} will cancel for all $k$ except where $\text{sgn}\{E_{n,\eta}(0,q,k)\} \neq \text{sgn}\{ E_{n,\eta}(\Delta,q,k)\}$. In the second line, we note that the $\Delta_z$ contribution cancels between inner and outer branches due to the fact that the product of the Cooper pair velocity of each branch, $v_\eta$, and the shift in momentum due to the Zeeman energy $\eta \frac{\Delta_z}{\hbar v_\eta}$ results in equal magnitude contributions with opposite signs (see SM~[\onlinecite{SI}] for details). Importantly, this means that the condition $j(q_0)=0$ is only satisfied by $q_0=0$ and there is no finite pairing momentum in the ground state. We will see below that this is not the case in the topological phase when $v_{\rm in}$ is absent. Although, analytically only valid at zero-temperature, we find this is a very good approximation also at finite temperature (see Fig.~\ref{topo}). 

\subsection{Diode efficiency in trivial phase.}
We define the diode efficiency as~\cite{Souto2022}
\begin{equation}
\delta=\frac{j_c^+-j_c^-}{j_c^++j_c^-},
\end{equation}
where $j_c^\pm$ are the critical currents to the left and right, such that $j_c^-=-\min\{j(q)\}$  and $j_c^+=\max\{j(q)\}$.  At zero temperature, we find that the extremal current densities occur precisely at the values of $q$, where the system becomes gapless as one of the energies $E_{n,\eta}(\Delta,q,k)$ in Eq.~\eqref{energy} becomes zero, i.e. when the condition $E_{n,\eta}(\Delta, q,k) >0$ for Eq.~\eqref{jq} is no longer satisfied for all $k$ (see SM~[\onlinecite{SI}] for details).~\cite{Legg2022} Note that $|\Delta_z|\geq\Delta$ results in a gapless system already for $q=0$, such that $\delta=0$ after this value and we assume throughout $0\leq |\Delta_z|<\Delta$.  From Eq.~\eqref{energy}, for each $n$ and $\eta$ we find that this occurs at the critical values 
\begin{equation}
\frac{1}{2} q_{\eta}^n= \eta \frac{ n  \Delta + \Delta_z }{\hbar  v_{\eta}}.
\end{equation}
Using Eq.~\eqref{jq}, the diode efficiency can be written as~\cite{Legg2022} 
\begin{equation}
\delta=\frac{q_c^{+}-q_c^{-}}{q_c^{+}+q_c^{-}}, \quad  q_c^{-}= -\max_{\eta} \{q_{\eta}^{-\eta} \}, \quad  q_c^{+}= \min_{\eta} \{q_{\eta}^{\eta} \},\label{qc}
\end{equation}
in other words the diode efficiency is set by the largest negative and smallest positive $q^n_\eta$. 
In particular, in the regime of small $|\Delta_z|$, both critical momenta  $q_c^\pm$ correspond to the same $\eta$ branch, namely that with the largest  $v_{\eta}$, and the diode efficiency grows linearly with Zeeman energy, $\Delta_z$.
For larger Zeeman strengths, $q_c^\pm$ correspond to different $\eta$ branches and the diode efficiency saturates to a constant. For instance, taking $v_{\rm out}>v_{\rm in}$, we find the diode efficiency is given by
\begin{equation}
\delta= \begin{cases}\frac{\Delta_z}{\Delta} & \left|\Delta_z\right|<\Delta \frac{v_{\text {out }}-v_{\text {in }}}{v_{\text {out }}+v_{\text {in }}} \\ \frac{\Delta_z}{\left|\Delta_z\right|}  \frac{v_{\text {out }}-v_{\text {in }}}{v_{\text {out }}+v_{\text {in }}} & \Delta \frac{v_{\text {out }}-v_{\text {in }}}{v_{\text {out }}+v_{\text {in }}} \leq\left|\Delta_z\right|<\Delta.  \end{cases}\label{delta}
\end{equation}
We emphasise that a finite diode efficiency is only possible when there is a difference in Fermi velocities $v_{\rm in}\neq v_{\rm out}$, because, otherwise, the contributions of two branches cancels each other even in the presence of a magnetic field. The resulting diode efficiency is shown in Fig.~\ref{delta1D} both using this analytic formula based on the linearised approximation and using the numerical maxima and minima of Eq.~\eqref{currentden} without the linearised approximation (see SM~[\onlinecite{SI}] for details). 

\subsection{Ground state Cooper pair momentum and diode efficiency in topological phase}

We now turn to the topological phase that can exist for a region of chemical potential when a parallel Zeeman field component, $\Delta_{x}$, is sufficiently large.~\cite{Laubscher2021} We note that for a SC diode effect to occur we still require a component of magnetic field such that $\Delta_z$ is also non-zero. This means that the outer branches, which are the only ones contributing to the current, are no longer Kramers partners. In general Eq.~\eqref{currentden} must now be calculated numerically, as was previously considered in Ref.~\citenum{Legg2022}. However, in this regime a finite pairing momentum $q_0$ can develop. In particular, if temperature is sufficiently low and $\Delta_x$ sufficiently large, such that the contributions from the inner Fermi points can be neglected, solving Eq.~\eqref{jq} for $j(q_0)=0$ in the absence of $v_{\rm in}$ contributions gives a finite pairing momentum $q_0\approx 2\Delta_{z}/\hbar v_{\rm out}$. As a result, deep in the topological phase, the critical current satisfies 
\begin{equation}
j_{c}^{\pm} = \frac{e v_{\rm out}}{\pi }  (q_{c}^{\pm}-q_{0}) \approx  \frac{2e \Delta}{\pi \hbar},
\end{equation} 
since $q_{c}^{\pm}=q^{\pm}_1=2(\Delta+\Delta_z)/\hbar v_{\rm out}$ because the inner branches no longer result in  critical momenta.
Note that $q_0$ is precisely the finite momentum due to the shift of the outer branches by the $\Delta_z$ component of the magnetic field, such that $q_0=0$ for $\Delta_z=0$.
As such, we see that the critical currents deep in the topological superconducting phase are once again (approximately) reciprocal and the diode efficiency $\delta\approx 0$ for sufficiently large $\Delta_{x}$. The full numerical results of $\delta$ and $q_0$ in this regime, for various temperatures $T$, are shown in Fig.~\ref{topo} and confirm this behavior (see SM~[\onlinecite{SI}] for details of numerics).

\section{Discussion}
We have investigated the relations between normal state nonreciprocal transport and the SC diode effect in proximitized nanowire systems. In both cases MCA can result in a difference in the inner and outer Fermi velocities and this is necessary to cause a finite diode effect. This difference in Fermi velocities generically arises either due to cubic SOI or due to a momentum dependent mass term in systems with SOI. We also found that the topology of the system strongly affects its nonreciprocal transport properties for both normal and superconducting systems. In the (almost) helical normal state we found that the rectification is generically enhanced due to the expected increase in scattering time. In contrast, the SC diode effect is substantially reduced due to the emergence of a finite Cooper pair momentum, $q_0$, deep in the topological phase. Our results indicate that systems with large nonreciprocal transport coefficients in the normal state can be prime candidates for SC diodes in proximitized systems since the difference in Fermi velocity of inner and outer pockets is key in both cases. However, despite their similarities in the trivial phase, we find that the impact of a topological phase can be strikingly different for normal state rectification in comparison to the SC diode effect. 

\begin{acknowledgments}
This work was supported by the Georg H. Endress Foundation and the Swiss National Science Foundation. This project received funding from the European Union’s Horizon 2020 research and innovation program (ERC Starting Grant, Grant No 757725). 
\end{acknowledgments}

\bibliographystyle{aapmrev4-1.bst}
\bibliography{diode.bib}
\end{document}


\preprint{AIP/123-QED}

\title{Supplemental material: \\Relations between normal state nonreciprocal transport and the superconducting diode effect in the trivial and topological phases}

\author{Georg Angehrn}
\affiliation{Department of Physics, University of Basel, Klingelbergstrasse 82, CH-4056 Basel, Switzerland}
\author{Henry F. Legg}
\affiliation{Department of Physics, University of Basel, Klingelbergstrasse 82, CH-4056 Basel, Switzerland}
\author{Daniel Loss}
\affiliation{Department of Physics, University of Basel, Klingelbergstrasse 82, CH-4056 Basel, Switzerland}
\author{Jelena Klinovaja}
\affiliation{Department of Physics, University of Basel, Klingelbergstrasse 82, CH-4056 Basel, Switzerland}

 \maketitle
In this supplemental material (SM) we derive the equation for the current  in a superconductor nanowire, i.e. Eq.~(10) of the main text. We also give the details of the numerics used for finite temperature calculations in Figs.~2 \& 3. Throughout this SM we follow closely a similar derivation in Ref.~\citenum{Legg2022}.
\section{Energy spectrum of superconducting nanowire}
We start with the general Hamiltonian, $h_k$, of the normal state of a nanowire [see Eq.~(1) of main text],
\begin{equation}
\label{hk}
h_{k} = \xi_{k}\sigma_{0} + (\alpha_{k} + \Delta_{z})\sigma_{3}+ \Delta_{x}\sigma_{1}-\mu. 
\end{equation}
We consider a pairing potential, $\Delta$, induced via proximity with a superconductor and in the presence of a current that results in a finite pairing momentum $q$, as in Eq.~(7) of the main text the Hamiltonian of such a setup is give by
\begin{align}
\mathcal{H}_{q}&=\frac{1}{2}\sum_k {\boldsymbol \psi}_k^\dagger \left(\begin{array}{cc}
h_{k+\frac{q}{2}} & -i \sigma_{y} \Delta \\
i \sigma_{y} \Delta & -h_{-k+\frac{q}{2}}^{*}
\end{array}\right) {\boldsymbol \psi}_k.
\end{align}
Here, as in Eq.~(7), ${\boldsymbol \psi}^\dagger_k$=($c^\dagger_{k+\frac{q}{2} \uparrow}$, $c^\dagger_{k+\frac{q}{2} \downarrow}$, $c_{-k+\frac{q}{2} \uparrow}$, $c_{-k+\frac{q}{2} \uparrow}$), where the  operator $c_{k \sigma}^\dagger$ ($c_{k \sigma}$) creates (annihilates) an electron with momentum $k$ and spin $\sigma$. 

Applying the Bogoliubov transformation for fermions, i.e. diagonalising in terms of new fermionic operators $\gamma^\dagger_{n,k,q}$ and $\gamma_{n,k,q}$ with $n=\pm$, we obtain
\begin{align}
\mathcal{H}_{q}&=\frac{1}{2}\sum_k \tilde{\boldsymbol \psi}_k^\dagger \left(\begin{array}{cccc}
\varepsilon_+(\Delta,q,k)+\varphi_+(\Delta,q,k) & 0&0&0 \\
0& \varepsilon_-(\Delta,q,k)+\varphi_-(\Delta,q,k)&0&0\\
0& 0&-\varepsilon_-(\Delta,q,k)+\varphi_-(\Delta,q,k)&0\\
0& 0&0&-\varepsilon_+(\Delta,q,k)+\varphi_+(\Delta,q,k)\\
\end{array}\right) \tilde{\boldsymbol \psi}_k,
\end{align}
where we defined
\begin{align}
\varepsilon_n(\Delta,q,k)=\sqrt{\Delta ^2+(\delta \alpha +n \bar{\xi} )^2},\\
\varphi_n(\Delta,q,k)=n(\bar{\alpha} +\Delta_z)+\delta \xi,
\end{align}
with $\bar{\xi}=\frac{1}{2}(\xi_{k+\frac{q}{2}}+\xi_{-k+\frac{q}{2}})$, $\delta\xi=\frac{1}{2}(\xi_{k+\frac{q}{2}}-\xi_{-k+\frac{q}{2}})$, $\bar{\alpha}=\frac{1}{2}(\alpha_{k+\frac{q}{2}}+\alpha_{-k+\frac{q}{2}})$, and $\delta\alpha=\frac{1}{2}(\alpha_{k+\frac{q}{2}}-\alpha_{-k+\frac{q}{2}})$ in terms of the general Hamiltonian in Eq.~\eqref{hk}.

To write this in a more manifestly particle-hole symmetric way we note that for $k \rightarrow -k$ we have: $\bar{\alpha}\to\bar{\alpha}$, $\delta \xi \to-\delta \xi$, $\bar{\xi} \to \bar{\xi}$, and $\delta{\alpha} \to -\delta{\alpha}$, such that $\varepsilon_n(\Delta,q,-k)=\varepsilon_{-n}(\Delta,q,k)$ and $\varphi_n(\Delta,q,-k)=-\varphi_{-n}(\Delta,q,k)$ and we can write
\begin{align}
\mathcal{H}_{q}&=\frac{1}{2}\sum_k \tilde{\boldsymbol \psi}_k^\dagger \left(\begin{array}{cccc}
E_+(\Delta,q,k)& 0&0&0 \\
0& E_-(\Delta,q,k)&0&0\\
0& 0&-E_+(\Delta,q,-k)&0\\
0& 0&0&-E_-(\Delta,q,-k)\\
\end{array}\right) \tilde{\boldsymbol \psi}_k,\label{phenergies}
\end{align}
where throughout $\tilde{\boldsymbol \psi}^\dagger_k=(\gamma^\dagger_{+,k,q}, \gamma^\dagger_{-,k,q}, \gamma_{+,-k,q}, \gamma_{-,-k,q}$) and we have defined the energies 
as
\begin{equation}
E_n(\Delta,q,k)=\varepsilon_n(\Delta,q,k)+\varphi_n(\Delta,q,k). \label{EnSC}
\end{equation}
Note that these energies can be either positive or negative and simply correspond to the upper $2\times2$ block of Eq.~\eqref{phenergies}.

These energies allow us to write the Hamiltonian as
\begin{equation}
\mathcal{H}_{q}=\sum_{k,n=\pm} \left[ E_n (\Delta,q,k)\gamma^\dagger_{n,k,q}\gamma_{n,k,q}-\frac{1}{2}E_n (\Delta,k,q) \right],
\end{equation}
 where the final term comes from the anticommutation relations of the fermionic operators $\gamma_{n,k,q}$.  Note that, for ease of analytic calculations, $E_n (\Delta,k,q)$ are defined without a modulus sign compared to Ref.~\citenum{Legg2022}, but we stress that the presence or not of the modulus sign does not affect any of our final results.

\section{Calculation of current}
\subsection{Current from the free energy}

The current in the nanowire can be obtained from the free energy density $\Omega(\Delta,q)$ which is given by~\cite{Kinnunen2018,Daido2022,Legg2022}
\begin{equation}
\Omega(\Delta,q)=-\frac{1}{2\pi }\int dk\sum_{n}  \frac{1}{\beta} \ln\left[\cosh(\beta E_n (\Delta,q,k)/2)\right],\label{freeE}
\end{equation}
where here we introduced $\beta=(k_{\rm B} T)^{-1}$ with $k_B$ the Boltzmann constant and $T$ temperature. Next, we can define the condensation energy density $F(q)=\Omega(\Delta,q)-\Omega(0,q)$. From $F(q)$, we can calculate the response to the current~\cite{Legg2022,Noah2022}
\begin{align}
j(q)&=\frac{2e}{\hbar} \partial_q F(q)=\frac{e}{\pi \hbar} \int dk \sum_{n} \big[\tanh(\beta E_n (0,q,k) /2) \partial_q  E_n (0,q,k)-\tanh(\beta E_n (\Delta,q,k) /2) \partial_q  E_n (\Delta,q,k)\big].\label{fulljq}
\end{align}
In the case of the linearised approximation, this gives Eq.~(11) of the main text. To obtain the finite temperature plots and/or plots for nonzero $\Delta_x$ in Figs.~2 \& 3, we numerically find the maximum and minimum values of $j(q)$ from Eq.~\eqref{fulljq} using the full energy spectrum as in Eq.~\eqref{EnSC}.
\subsection{Current at zero temperature}

At zero temperature $\tanh(\beta x)\rightarrow \Theta(x)-\Theta(-x)$, where $\Theta(x)$ is the Heaviside function. This gives a current 
\begin{equation}
j(q)=\frac{e}{\pi \hbar} \int dk \sum_{n} \big\{[\Theta(E_n (0,q,k))-\Theta(-E_n (0,q,k))] \partial_q  E_n (0,q,k)-[\Theta(E_n (\Delta,q,k))-\Theta(-E_n (\Delta,q,k))] \partial_q  E_n (\Delta,q,k) \big\}, \label{zeroTj}
\end{equation}
where we made use of $\Theta (x/2)=\Theta(x)$.

\section{Current from linearised spectrum}

As in Eq.~(8) of the main text, we now linearise the normal state spectrum about the Fermi energy
\begin{equation}
h_{\eta k}= \hbar v_\eta (-\eta  \sigma_3 k - k_\eta^0)+\Delta_z \sigma_3,
\end{equation}
where $\eta=+1 (-1)$ indicates the outer (inner) Fermi points, such that the $k_{-1}^0$ ($k_{1}^0$) is the Fermi wave number of the inner (outer) point of the $+$ band at zero magnetic field and velocities $v_{-1}=v_{\rm in}$ and $v_{1}=v_{\rm out}$ are as defined in the main text. 

\begin{figure}[h]
\includegraphics[width=1\linewidth]{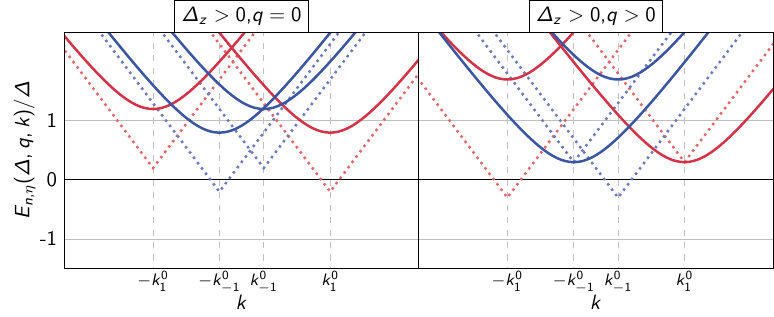}
\caption{{\it The energy spectrum}: The linearised spectrum $E_{n,\eta} (\Delta,k,q)$[$E_{n,\eta} (0,k,q)$] is shown by solid [dashed] lines. The regions where ${\rm sgn} \{ E_{n,\eta} (0,k,q)\}\neq {\rm sgn} \{ E_{n,\eta} (\Delta ,k,q)\}$ are the only ones that contribute to $j(q)$ in Eq.~\eqref{jqlinear}. Blue (red) indicates inner (outer) branches corresponding to $\eta=-1$ ($\eta=1$) and Fermi momentum $k^0_{-1}$ ($k^0_{1}$). Note that for finite $q$, one band from each branch moves up in energy for all $k$ and the other moves down.}
\label{fig1}
\end{figure}

As such, for this linearised spectrum, $\xi_{k}=-\hbar v_\eta k^0_\eta$ is a constant and $\alpha_k=-\hbar v_\eta \eta k$.  Using the definitions above, this gives $\bar{\xi}=-\hbar v_\eta k^0_\eta$, $\delta\xi=0$, $\bar{\alpha}=-\hbar v_\eta \eta \frac{q}{2}$, and $\delta\alpha=-\hbar v_\eta \eta k$. Leading to
\begin{align}
\varepsilon_{n,\eta}(\Delta,q,k)=\sqrt{\Delta ^2+\hbar^2 v_\eta^2 ( n \eta k +  k^0_\eta)^2}
\end{align}
and
\begin{align}
\varphi_{n,\eta}(\Delta,q,k)=n(-\hbar v_\eta \eta q/2 +\Delta_z).
\end{align}
Such that the superconductor energy spectrum corresponding to each pair of bands, $\eta$, defined in  Eq.~\eqref{phenergies} (see also Fig.~\ref{fig1}) is written as
\begin{equation}
E_{n,\eta}(\Delta,q,k)=\varepsilon_{n,\eta}(\Delta,q,k)+\varphi_{n,\eta}(\Delta,q,k)=\sqrt{\Delta ^2+\hbar^2 v_\eta^2 ( n \eta k +k^0_\eta)^2}+n(-\hbar v_\eta \eta q/2 +\Delta_z).
\end{equation}
In particular, we find that
\begin{equation}
\partial_q E_{n,\eta}(\Delta,q,k)=-n \hbar v_\eta \eta/2 ,
\end{equation}
in other words this derivative is independent of both $k$, $q$, and $\Delta$. This allows us to simplify Eq.~\eqref{zeroTj} for the current at zero-temperature, such that
\begin{equation}
\begin{aligned}
j(q)&=\frac{-e}{2\pi } \int dk \sum_{n,\eta} n  v_\eta \eta \big\{[\Theta(E_{n,\eta} (0,q,k))-\Theta(E_{n,\eta} (\Delta,q,k))]+[\Theta(-E_{n,\eta} (\Delta,q,k))-\Theta(-E_{n,\eta} (0,q,k))]  \big\}. \label{zeroJ}
\end{aligned}
\end{equation}
In particular, if we consider values of $q$ for which $E_{n,\eta} (\Delta,q,k)>0$ for all $k$ ($\Delta -| \Delta_z| > |\hbar v_\eta q/2| $), then 
\begin{equation}
\begin{aligned}
j(q)=\frac{-e}{2\pi } \sum_{n,\eta} n v_\eta \eta  \int dk \;\;\bigg({\rm sgn}\{E_{n,\eta} (0,q,k)\}-1\bigg).\label{jqlinear}
\end{aligned}
\end{equation}
In other words the integrand is only non-zero if $ \hbar v_\eta |( n \eta k + k^0_\eta)| + n(-\hbar v_\eta \eta q/2 +\Delta_z)< 0$.
For each branch, inner and outer, only a single region satisfies this equality (see Fig.~\ref{fig1}). In particular, the extremal momenta of this region, $k_1$ and $k_2$, satisfy $\hbar v_\eta( n \eta k_1+ k^0_\eta) = n (-\hbar v_\eta \eta q/2 +\Delta_z)<0$ and $-\hbar v_\eta(n \eta k_2+ k^0_\eta) = n (-\hbar v_\eta \eta q/2 +\Delta_z) <0$, such that we get to $\eta n (k_2-k_1)=n \eta q - \frac{2 n \Delta_z}{\hbar v_\eta}\geq0$.
The integral over $k$ yields $-2\left(\eta n q - \frac{2 n \Delta_z}{\hbar v_\eta}\right)\Theta\left(\eta n q - \frac{2 n \Delta_z}{\hbar v_\eta}\right)$. Inserting this integral into Eq.~\eqref{jqlinear} for the current gives
\begin{equation}
\begin{aligned}
j(q)=  \frac{e}{\pi } \sum_{\eta} v_\eta \left(q - \frac{2\eta \Delta_z}{\hbar v_\eta } \right)= \frac{e}{\pi }   q (v_{\rm in}+v_{\rm out}),
\end{aligned}
\end{equation}
as is found in the main text. Note that the Zeeman energy drops out due to the cancellation of velocity factors.

Finally, we note that in Eq.~\eqref{zeroJ}, if we allow for some $k$ where $E_{n,\eta} (\Delta,q,k)<0$ then these regions will reduce the current compared to the maximal value allowed when $E_{n,\eta} (\Delta,q,k)>0$ is true for all $k$ (see the main text).

\section{Critical momenta}
Here, we give further details of the diode efficiency $\delta$, as in Eq.~(15) of the main text. For small $\Delta_z$, the critical momenta $q_c^{\pm}$ correspond to the same branch, $\eta$, with the largest Fermi-velocity $v_\eta$. Assuming $v_{\rm out}>v_{\rm in}$, we obtain $q_c^{\pm}=\frac{\Delta \pm \Delta_z}{\hbar v_{\rm out}}$ and $\delta=\frac{q_c^{+}-q_c^{-}}{q_c^{+}+q_c^{-}}=\frac{\Delta_z}{\Delta}$. On the other hand, for $\Delta \frac{v_{\text {out }}-v_{\text {in }}}{v_{\text {out }}+v_{\text {in }}} \leq \left|\Delta_z\right| < \Delta$, opposite $\eta$ branches are responsible for the critical momenta $q_c^{\pm}$. In particular, taking $\Delta_z>0$, we obtain $q_c^{+}=\frac{\Delta - \Delta_z}{\hbar v_{\rm in}}$ and $q_c^{-}=\frac{\Delta - \Delta_z}{\hbar v_{\rm out}}$, which gives $\delta=\frac{v_{\text {out }}-v_{\text {in }}}{v_{\text {out }}+v_{\text {in }}}$. Conversely, for $\Delta_z <0$, in this regime, the result is $q_c^{+}=\frac{\Delta + \Delta_z}{\hbar v_{\rm out}}$ and $q_c^{-}=\frac{\Delta + \Delta_z}{\hbar v_{\rm in}}$, such that we obtain $\delta=-\frac{v_{\text {out }}-v_{\text {in }}}{v_{\text {out }}+v_{\text {in }}}$.

\bibliographystyle{apsrev4-1}
\bibliography{diode.bib}